\documentclass[prl,aps,a4paper,10pt,twocolumn,superscriptaddress,amsmath,amssymb,showpacs]{revtex4}
\usepackage{bm}
\usepackage{epsfig}
\usepackage{amsmath}

\renewcommand{\v}[1]{{\bf #1}}
\newcommand{\bpm}{\begin{pmatrix}}
\newcommand{\epm}{\end{pmatrix}}
\newcommand{\ba}{\begin{eqnarray}}
\newcommand{\ea}{\end{eqnarray}}
\newcommand{\nn}{\nonumber \\}

\begin{document}

\title{Detecting Chiral Orbital Angular Momentum by Circular Dichroism ARPES}
\author{Jin-Hong Park}
\affiliation{Department of Physics and BK21 Physics Research
Division, Sungkyunkwan University, Suwon 440-746, Korea}
\author{Choong H. Kim} \affiliation{Department of
Physics and Astronomy, Seoul National University, Seoul 151-742,
Korea}
\author{Jun Won Rhim}
\affiliation{School of Physics, Korea Institute for Advanced Study,
Seoul 130-722, Korea}
\author{Jung Hoon Han}
\email[Electronic address:$~~$]{hanjh@skku.edu}
\affiliation{Department of Physics and BK21 Physics Research
Division, Sungkyunkwan University, Suwon 440-746, Korea}

\date{\today}

\begin{abstract} We show, by way of tight-binding and first-principles calculations,
that a one-to-one correspondence between electron's crystal momentum
$\v k$ and non-zero orbital angular momentum (OAM) is a generic
feature of surface bands. The OAM forms a chiral structure in
momentum space much as its spin counterpart in Rashba model does, as
a consequence of the inherent inversion symmetry breaking at the
surface but not of spin-orbit interaction. Circular dichroism (CD)
angle-resolved photoemission (ARPES) experiment is an efficient way
to detect this new order, and we derive formulas explicitly relating
the CD-ARPES signal to the existence of OAM in the band structure.
The cases of degenerate $p$- and $d$-orbital bands are considered.
\end{abstract}

\maketitle

Electron spins are quenched in ordinary crystalline solids in the
sense that each crystal momentum $\v k$ comes in degenerate pairs of
spin-up and spin-down electrons. Such degeneracy is lifted in an
interesting manner for the so-called Rashba system\cite{rashba}, in
which the given momentum state in the band has only one spin state
associated with it. Examples of Rashba-split bands are many by
now\cite{bihlmayer-review}. On symmetry grounds, as Rashba originally
argued, the inherent inversion symmetry breaking (ISB) at the surface
termination allows an interaction term, the Rashba term, of the form
$H_\mathrm{R} = \lambda_\mathrm{R} \hat{z} \cdot (\v k \times \bm
\sigma )$ involving the coupling of the electron's spin operator $\bm
\sigma/2$ and its momentum $\v k$ (We set $\hbar \equiv 1$). The
chiral spin angular momentum (SAM) structure in momentum space
follows as a direct consequence of the Rashba Hamiltonian
$H_\mathrm{R}$\cite{SARPES,kimura}.

One must note, however, that other physical quantities having the
character of angular momentum can take the place of $\bm \sigma$ in
the Rashba Hamiltonian. For instance, in bands where the orbital
angular momentum (OAM) remains unquenched, one can equally well
consider a coupling of $\v k$ to OAM operator, $\v L$. In this paper,
we expand this observation to show that chiral OAM in one-to-one
correspondence with the electron's linear momentum is indeed a
general consequence of ISB for the surface bands. Unlike the chiral
SAM structure which typically occurs in materials with strong
spin-orbit interaction (SOI)\cite{bihlmayer-review}, we show that
chiral OAM occurs even for materials with no SOI.

We begin by providing a simple, tight-binding (TB) model Hamiltonian
in two dimensions where the above-mentioned chiral OAM structure
emerges. A three-fold degenerate $p$-orbital system is considered,
while the spin degrees of freedom is suppressed in order to emphasize
the notion of chiral OAM emerging from ISB alone without the
requirement of SOI. The TB model is the triangular lattice version of
the square one originally considered by Petersen and
Hedeg{\aa}rd\cite{petersen} in their study of Rashba phenomena in
$p$-orbital bands. By suppressing spin degrees of freedom, we do away
with the notion of conventional Rashba spin splitting. The idea of
ISB-induced chiral OAM was not considered in Ref.
\onlinecite{petersen} or in any other literature we are aware
of\cite{ganichev-OAM}.

We introduce two Slater-Koster parameters $V_1$ and $V_2$ for
$\sigma$- and $\pi$-bonding amplitudes, respectively, and a third
one, $\gamma$, representing the degree of ISB\cite{petersen}. We will
write $N$ for the number of sites in the lattice, and $|i, \lambda
\rangle$ for the localized Wannier orbitals $(\lambda=p_x, p_y, p_z$)
at the atomic site $\v r_i$. Then the tight-binding Hamiltonian for
the triangular lattice (essentially the same result obtains for
square lattice) near the $\Gamma$-point ($\v k = 0$) in the
momentum-space basis $|\v k , \lambda \rangle = N^{-1/2} \sum_i e^{i
\v k \cdot \v r_i} | i, \lambda \rangle$ becomes

\ba H_{\v k}=\bpm
\alpha k_x^2+\beta k_y^2&(\alpha-\beta)k_x k_y&-i\frac{3}{2}\gamma k_x \\
(\alpha-\beta)k_xk_y &\alpha k_y^2+\beta k_x^2 & -i\frac{3}{2}\gamma
k_y\\i\frac{3}{2}\gamma k_x&i\frac{3}{2}\gamma
k_y&4(\alpha-\beta)-\frac{3}{2}V_2k^2\epm . \label{eq:OAM-model}\ea
Here, $\alpha=3(3V_1-V_2)/8$, $\beta=3(V_1-3V_2)/8$ and
$k^2=k_x^2+k_y^2$. The lattice constant is taken to be unity. To
diagonalize $H_{\v k}$, it is convenient to choose a new set of
basis vectors

\ba |\v k, \mathrm{I} \rangle &=& (k_y /k) |\v k, p_x \rangle -(k_x
/k) |\v k, p_y \rangle, \nn
|\v k, \mathrm{II}\rangle &=& (k_x /k) |\v k, p_x \rangle +(k_y /k)
|\v k, p_y \rangle, \nn
|\v k, \mathrm{III}\rangle &=& e^{-i\phi_{\v k}} |\v k, p_z \rangle,
\label{eq:8} \ea
$k =|\v k |$, $e^{i\phi_{\v k}} = (k_x + ik_y)/k$. The state $|\v k,
\mathrm{I}\rangle$ remains decoupled at energy $E_{1,\v k} = 3 V_2 -
3 V_1 +3(V_1-3 V_2) \v k^2 /8$, while $|\v k, \mathrm{II}\rangle$ and
$|\v k, \mathrm{III}\rangle$ combine to form eigenstates which, to
leading order of $\gamma /\Delta$, are

\ba |\v k , 2 \rangle &\simeq&  |\v k, \mathrm{II}\rangle - {i \gamma
(k_x - ik_y) \over 2\Delta}|\v k, \mathrm{III}\rangle, \nn
|\v k, 3 \rangle &\simeq &|\v k, \mathrm{III} \rangle - {i\gamma (k_x
+ ik_y) \over 2\Delta}|\v k, \mathrm{II} \rangle , \label{eq:10}\ea
with energies $E_{2, \v k} \simeq 3(V_2-V_1) +3 ((3V_1 - V_2 )/8
-\gamma^2/4\Delta) \v k^2$ and $E_{3, \v k} \simeq 6V_2 +
(3\gamma^2/4\Delta-3V_2/2) \v k^2$. The OAM operator is given by the
sum $\v L = (1/N) \sum_i \v L_i$ where each $\v L_i$ acts on the
Wannier state $|i , \lambda\rangle$ as the usual $L=1$ angular
momentum operator. The two bands obtained above carry nonzero, chiral
OAM as claimed ($L^+ = L^x + i L^y$):

\ba \langle \v k , 2 |L^+|  \v k , 2 \rangle \!=\! {i \gamma \over
\Delta } (k_x \!+\! i k_y ) \!=\! - \langle \v k , 3 |L^+|\v k , 3
\rangle . \label{eq:perturbative-OAM}\ea
The ``helicities" of the chiral OAMs are opposite between the bands,
with their magnitudes vanish linearly with $|\v k|$ and the ISB
parameter $\gamma$. In contrast the Rashba model displays perfect
spin polarization irrespective of the wave vector or the strength of
the Rashba parameter $\lambda_\mathrm{R}$.

\begin{figure}[ht]
\includegraphics[width=80mm]{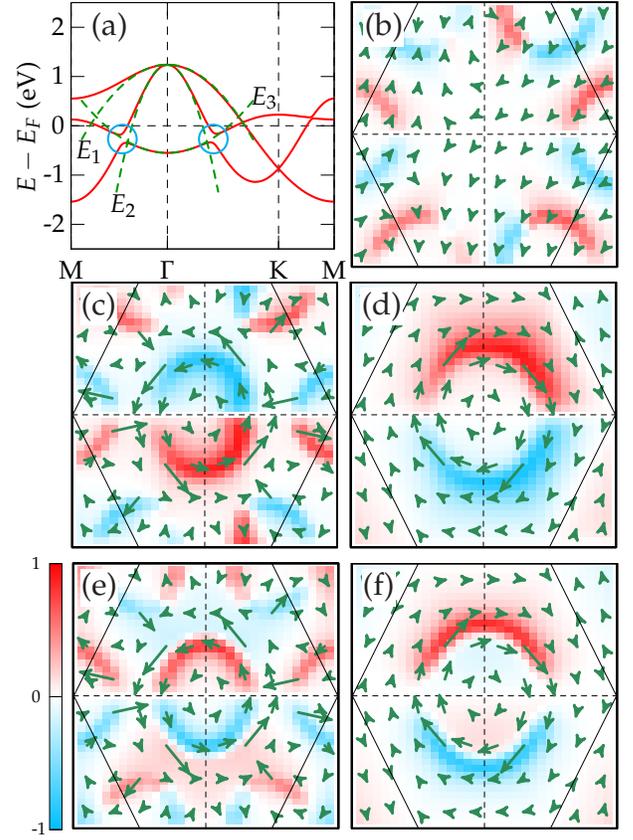}
\caption{OAM and CD from first-principles and tight-binding
calculations of Bi monolayer without SOI. (a) LDA band structure for
Bi monolayer with SOI turned off. Perpendicular electric field of
3V/\AA~was imposed externally. Three dashed curves represent the
tight-binding energy dispersions around the $\Gamma$ point. (b)-(d)
OAM vectors (green arrows) and NCD signals (color backgrounds) for
the three bands, $E_1$(b), $E_2$(c), and $E_3$(d) over the whole
Brillouin zone marked by a solid hexagon. Largest OAM has magnitude
$\approx 1\hbar$ for bands $E_2$ and $E_3$. NCD is calculated with
$k_{F,z} = 2.27$\AA$^{-1}$. (e)-(f) NCD calculated with $k_{F,z} = 0$
in Eq. (\ref{eq:9}) for $E_2$(e) and $E_3$(f) bands. The opposite
color assignments between (c) and (e) is a consequence of photon
energy dependence of the scattering intensity.}
\label{fig:LDA-CD-woSO}
\end{figure}

A second check on the existence of chiral OAM is performed by
employing the first-principles local-density approximation (LDA)
calculation for a Bi single layer forming a triangular lattice. The
choice is inspired by Bi being a proto-typical $p$-orbital band
material. An external electric field of 3V/\AA~ perpendicular to the
layer was imposed by hand to mimic the surface potential gradient
without having the complication of dealing with the bulk states. We
also performed calculations for the physically more realistic case of
Bi bi-layer with perpendicular electric field\cite{liu}, with results
that are entirely in accord with the statements made below for the
single-layer case regarding the emergence of chiral OAM. To emphasize
the relevance of ISB we again chose to investigate the
spin-degenerate case by turning off SOI in the LDA calculation. The
resulting electronic structure for spinless case consisting of three
$p$-orbital-derived bands is shown in Fig.~\ref{fig:LDA-CD-woSO}(a).
As the external electric field is turned on, a level repulsion
between the middle ($E_2$ in Fig. 1(a)) and the bottom ($E_3$ in Fig.
1(a)) band occurs as indicated by circles in
Fig.~\ref{fig:LDA-CD-woSO}(a). These two bands exhibit the chiral OAM
patterns with the maximum OAM vector $|\langle \bm L \rangle |
\approx 0.96\hbar$ as shown in Fig.~\ref{fig:LDA-CD-woSO}(c) and
1(d), while the third one, shown in Fig. 1(b), carries much less OAM
around the $\Gamma$ point. The OAM chiralities of the two bands are
opposite, in accordance with the previous TB analysis. An excellent
fit of the LDA band structure near the $\Gamma$ point was possible
with the TB parameters $V_1=-0.725$ eV, $V_2=-0.11$ eV and
$\gamma=0.2623$ eV (Fig. 1(a)). The OAM magnitude is seen to decrease
continuously upon approaching the $\Gamma$ point in the LDA
calculation (Fig. 1(c) and 1(d)) as predicted by the TB calculation,
Eq. (\ref{eq:perturbative-OAM}).

Having established theoretically the existence of chiral OAM in
generic inversion-asymmetric bands by a number of methods, we turn to
the question of its detection. Spin- and angle-resolved photoemission
spectroscopy (SARPES) has served to identify the chiral spin
structure of the surface bands in the past\cite{SARPES,kimura}. A
similar chiral structure for OAM as demonstrated here cannot,
however, be detected by the same probe since chiral OAM exists even
when SOI is very weak and spin degeneracy is nearly perfect. Given
the potential prevalence of chiral OAM in ordinary, \textit{i.e.}
weak SOI, surface bands, it is essential to establish detection tools
for this new degree of freedom in band materials and develop some
theory of it.

Circular dichroism (CD) refers to phenomena in which the physical
response of a system to probing light depends systematically on the
light polarization being left-circularly-polarized (LCP) or
right-circularly-polarized (RCP). In ARPES, CD manifests itself as
different scattering intensities of the photo-electrons depending on
the helicity of incident light being RCP or LCP. The CD-ARPES signal
can be quantified through the normalized CD (NCD) defined as

\ba D(\v k) &=& {\sum_\sigma  \bigl(  I^{\rm{RCP}}_\sigma (\v k) -
I^{\rm{LCP}}_\sigma (\v k) \bigr) \over \sum_\sigma \bigl(
I^{\rm{RCP}}_\sigma (\v k) + I^{\rm{LCP}}_\sigma (\v k)\bigr)} .
\label{eq:CD-formula}\ea
The sum over the final state spin $\sigma$ reflects the
spin-integrated nature of the detection scheme. Spin index is
restored in the following derivation of the NCD formula as it will
apply equally well to strong-SOI materials with mixed spins.

The initial state $|\mathrm{I}\rangle$ is the Bloch state of
momentum $\v k$ constructed as $ |\v k, \bm m \rangle = N^{-1/2}
\sum_i e^{i \v k \cdot \v r_i} | i, \bm m \rangle$. The Wannier
state $|i, \bm m \rangle$ at site $\v r_i$ is given by

\ba | i , \bm m \rangle  = \sum_{\lambda, \sigma} m_{\lambda,
\sigma} (\v k) |i, \lambda, \sigma \rangle ,
\label{eq:initial-state}\ea
as a linear combination of constituent atomic orbitals labeled by
$\lambda$, and spin $\sigma$, with $\v k$-dependent coefficients
$m_{\lambda, \sigma} (\v k)$. Usual plane wave forms are assumed for
the final state: $\psi_F \sim e^{i\v k_F \cdot \v r}$.

The transition amplitude into the final state of spin $\sigma$ is
evaluated as

\ba && \langle F, \sigma | \v p \cdot \v A | I \rangle \sim \langle
F, \sigma | \v r \cdot \v A |I \rangle \nn
&& ~~~~ \sim \sum_{i, \lambda} m_{\lambda, \sigma} (\v k)\langle F |
\v r \cdot \v A | i, \lambda \rangle e^{i \v k \cdot \v r_i}. \ea
Given the localized nature of the Wannier state, it is useful to
re-write $\v r \cdot \v A = (\v r-\v r_i ) \cdot \v A + \v r_i \cdot
\v A$. One immediately finds that $\langle F | \v r_i \cdot \v A |I
\rangle = (\v r_i \cdot \v A) \langle F| I \rangle$ is zero from the
presumed orthogonality of the initial and the final states. To
proceed further, we treat the cases of $p$-orbital and $d$-orbital
bands separately as they require somewhat different strategies for
evaluation of the NCD formula. For $p$-orbitals the Wannier states
are assigned the hydrogenic wave function $\langle \v r | i, \lambda
\rangle \sim (\v r - \v r_i )_\lambda f(|\v r - \v r_i |)$ ($\lambda
= x,y,z$) and with these we find the transition amplitude becomes

\ba && \sum_{i} e^{i (\v k - \v k_F ) \cdot \v r_i }\left(
\sum_\lambda m_{\lambda, \sigma} (\v k) \int ~ e^{-i \v k_F \cdot \v
r} [ \v r \cdot \v A ] x_\lambda f(| \v r | ) \right) \nn
&& \propto \sum_\lambda m_{\lambda, \sigma} (\v k) \int ~ e^{-i \v
k_F \cdot \v r} [ \v r \cdot \v A ] x_\lambda f(| \v r | ).
\label{eq:7} \ea
The sum $\sum_i e^{i (\v k - \v k_F ) \cdot \v r_i } = \delta (\v
k_F^\parallel - \v k + \v G)$ simply yields in-plane momentum
conservation up to the reciprocal lattice vector $\v G$. Making use
of the fact that $f(|\v r|)$ depends only on the radial distance, one
can re-write the second line of Eq. (\ref{eq:7}) as

\ba - A_\alpha m_{\beta, \sigma } \partial_\alpha
\partial_\beta \int e^{-i \v k_F \cdot \v r} f (| \v r | ) d^3 \v r  = -
\v A \cdot \bm \nabla_{\v k_F} g_\sigma  (\v k_F ) ,\ea
where we have introduced

\ba f(| \v k_F | ) &=& \int e^{-i\v k_F \cdot \v r} f(|\v r|) d^3 \v
r ,\nn
g_\sigma (\v k_F ) &=& \bm m_\sigma \cdot \bm \nabla_{\v k_F} f (|\v
k_F |), \label{eq:f-and-g}\ea
and $\bm m_\sigma = (m_{x,\sigma}, m_{y,\sigma}, m_{z,\sigma})$. One
finds the average OAM $\langle \v L \rangle$ for the $p$-orbital
Bloch states $|\v k, \bm m\rangle$ expressed simply in terms of $\bm
m_\sigma$:

\ba   \langle \v k, \bm m | \v L | \v k , \bm m \rangle \equiv
\langle \v L \rangle = i \sum_\sigma \bm m_\sigma (\v k) \times \bm
m_\sigma^* (\v k) . \label{eq:OAM-average-for-p}\ea
Of great importance is the fact that nonzero OAM is possible even
with spin degeneracy, $\bm m_\sigma = \bm m$, where one would obtain
$\langle \v L \rangle = 2i \bm m \times \bm m^*$. We can now proceed
to show that NCD is fundamentally related to $\langle \v L \rangle$.
Equations (\ref{eq:7}) and (\ref{eq:f-and-g}) allow the NCD formula
(\ref{eq:CD-formula}) to be re-cast in the compact, suggestive form:

\ba D(\v k) \!  = \!  {(\v A\times \v A^*  ) \cdot \sum_\sigma \bm
\nabla g_\sigma \times \bm \nabla g_\sigma^* \over \sum_\sigma \left[
(\v A \! \cdot \! \bm \nabla g_\sigma ) (\v A^* \! \cdot \! \bm
\nabla g_\sigma^* ) \!+\!  (\v A^* \! \cdot \! \bm \nabla g_\sigma
)(\v A \! \cdot \! \bm \nabla g_\sigma^* )  \right] } .
\label{eq:9}\ea
Explicit reference to $\v k_F$ in the derivatives has been dropped
from above.  The vector potentials are $\v A = (\bm \varepsilon_1 + i
\bm \varepsilon_2 )/\sqrt{2}$ for RCP and $\v A^*$ for LCP, with $\v
A \times \v A^* = -i \bm \varepsilon_1 \times \bm \varepsilon_2 = -i
\hat{k}_\mathrm{ph}$ giving the photon direction. The remaining task
is to evaluate the quantity $\sum_\sigma \bm \nabla g_\sigma \times
\bm \nabla g_\sigma^*$ which governs the CD response of the given
initial state. The denominator, by definition, is always positive and
plays a minor role in characterizing the OAM structure.

For $p$-orbitals, inserting $g_\sigma = \bm m_\sigma \cdot \bm \nabla
f$ gives

\ba \sum_\sigma  \bm \nabla g_\sigma \times \bm \nabla g^*_\sigma &=&
{1\over 2} \varepsilon^{\alpha \beta \gamma} ( \bm m_\sigma \times
\bm m_\sigma^* )_\alpha \bm \nabla
\partial_\beta f \times \bm \nabla \partial_\gamma f \nn
&=& -{i\over 2} \varepsilon^{\alpha \beta \gamma} \langle L_\alpha
\rangle \bm \nabla \partial_\beta f \times \bm \nabla
\partial_\gamma f , \label{eq:simple-p}\ea
where Eq. (\ref{eq:OAM-average-for-p}) has been adopted in the
second line. Clearly, this is proportional to the components of OAM
in the initial state. The remaining ``form factor" $\bm \nabla
\partial_\beta f \times \bm \nabla \partial_\gamma f$ depends on $\v k_F$,
which in turn depends on the incoming photon energy. Both $\langle
\v L \rangle$ and the form factors can be obtained by faithful LDA
calculations of the wave functions. Analytically, choosing $f(r)
\sim e^{-r/a}$, for instance, yields $f(k_F ) = 1/(1+k_F^2 a^2)^2$
and the form factors can be worked out.

Next we turn to the case of degenerate $d$-orbital bands. Now
$g_\sigma = ( \bm m_\sigma \cdot \v D ) f$ involves the inner
product between the five-dimensional coefficient vector $\bm
m_\sigma$ and the corresponding differential operators $D_\alpha$
matching the given orbital basis. Examples are $D_{xy} = \partial_x
\partial_y$, $D_{3z^2 - r^2}  = (2\partial_z^2 - \partial_x^2 -
\partial_y^2)/2\sqrt{3}$, etc. Although the dichroism formula (\ref{eq:9})
still applies, we are no longer able to transform its numerator to a
simple form like Eq. (\ref{eq:simple-p}). Recall that in realistic
ARPES experiment the incident photons carry energies of several tens
of eV, delivering however at most an eV of energy to the occupied
electrons\cite{shen-review}. After subtracting what amounts to the
surface potential energy barrier, there is still a lot of energy
imparted to the final photo-electron, whose energy is typically in
excess of 10 eV. Due to in-plane momentum conservation, the in-plane
component of photo-electron momentum can carry only a small fraction
of this energy, which means most of the kinetic energy is contained
in the $z$-component, $(k_{F,z} )^2/2m$. As a result, the typical
situation in ARPES experiment is the one in which $k_{F,z}$
dominates over the planar components, and $k_{F,z}a$ is rather
larger than unity, $a$ being the radius of the orbital wave
function.

In computing $g_\sigma = (\bm m_\sigma \cdot \v D) f$, therefore,
the orbitals containing at least one power of $z$ will be dominant
over those that contain none, due to extra powers of $k_F^z a$
produced by the differentiation. The reasoning reduces the number of
relevant orbitals from five to three, i.e. $d_{zx}$, $d_{yz}$, and
$d_{3z^2 - r^2}$. This is irrespective of particular crystal field
symmetries of the $d$-orbitals and is rather dictated by the
experimental conditions of ARPES. We also introduce the ``partial
OAM" obtained by assuming only the three, labeled $1 \equiv zx$, $2
\equiv yz$, and $3 \equiv 3z^2 - r^2$, out of the five $d$-orbitals
contribute to the wave function: $\langle L_x \rangle' = \sqrt{3} i
\sum_\sigma \Bigl(m_{2,\sigma} m_{3,\sigma}^* -m_{2,\sigma}^*
m_{3,\sigma}\Bigr), \langle L_y \rangle' = \sqrt{3} i \sum_\sigma
\Bigl(m_{3,\sigma} m_{1,\sigma}^* -m_{3,\sigma}^*
m_{1,\sigma}\Bigr), \langle L_z \rangle' = i \sum_{\sigma} \Bigl(
m_{1,\sigma} m_{2,\sigma}^* - m_{1,\sigma}^* m_{2,\sigma} \Bigr)$.
The prime is a reminder that contributions average OAM from $d_{xy}$
and $d_{x^2-y^2}$ are missing. Following the same calculation
procedure as in the $p$-orbital case, we find the CD formula for
$d$-orbitals valid for $k_{F,z}a \gg 1$,

\begin{widetext}
\ba D_d (\v k) &=& {\hat{ k}_\mathrm{ph} \cdot (2 \langle L_x
\rangle' \hat{x} + 2 \langle L_y \rangle' \hat{y}- \langle L_z
\rangle' \hat{z} )\over \sum_\sigma [A_x m_{1, \sigma} + A_y m_{2,
\sigma} - 2\sqrt{3} A_z m_{3, \sigma}] [(A_x)^* m_{1, \sigma}^* +
(A_y)^* m_{2, \sigma}^* - 2\sqrt{3} (A_z)^* m_{3, \sigma}^*] \!+\!
(\v A \leftrightarrow \v A^* ) }. \label{eq:large-kf-CD}\ea
The OAM shown in the numerator refers to those in partial OAM. Again,
non-trivial CD-ARPES signal is predicated on the existence of OAM. In
the same limit, $k_{F,z}a\gg 1$, the $p$-orbital formula simplifies
to

\ba D_p (\v k) = { 5 \hat{ k}_\mathrm{ph} \cdot \langle \v L \rangle
\! -\! 6 (\hat{k}_\mathrm{ph} \cdot \hat{k}_F) (\hat{k}_F \cdot
\langle \v L \rangle)  \over \sum_\sigma [  6 (\hat{k}_F \cdot \v A)
(\hat{k}_F \cdot \bm m_\sigma ) - (\bm m_\sigma \cdot \v A) ] [ 6
(\hat{k}_F \cdot \v A^* ) ( \hat{k}_F \cdot \bm m_\sigma^* ) - (\bm
m_\sigma^* \cdot \v A^* ) ] + (\v A \leftrightarrow \v A^* ) } .
\label{eq:p-orbital-CD}\ea
\end{widetext}
The case of $t_{2g}$-bands involving the three orbitals, $yz \equiv
1$, $zx \equiv 2$, and $xy \equiv 3$ can be considered as well. The
factor $\sum_\sigma \bm \nabla g_\sigma \times \bm \nabla g_\sigma^*$
in Eq. (\ref{eq:9}) becomes

\ba {1 \over 2}\varepsilon^{\alpha\beta\gamma} \Bigl(\sum_{\sigma}
\bm m_\sigma \times \bm m_\sigma^* \Bigr)_\alpha \bm \nabla D_\beta f
\times \bm \nabla D_\gamma f, \label{eq:d-expand-eq}\ea
where $\v D = (D_{yz}, D_{zx}, D_{xy})$. As in the $p$-orbital case,
we obtain a simple relation for the OAM average of the given Bloch
state in the $t_{2g}$-band: $\langle \v L \rangle = -i \sum_\sigma
\bm m_\sigma \times \bm m^*_\sigma$. (Note the opposite sign compared
to the $p$-orbital result in Eq. (\ref{eq:OAM-average-for-p}).)
Again, non-trivial OAM is responsible for NCD for $t_{2g}$-bands. The
actual task of evaluating the form factors leads to cumbersome
expressions. For practical applications, it is better to evaluate
them numerically using LDA-obtained wave functions.

The OAM of the Bi bands shown in Fig. \ref{fig:LDA-CD-woSO}(a) is
used to obtain $D(\v k)$ according to Eqs. (\ref{eq:9}) and
(\ref{eq:simple-p}). Exponential function $f(r)\sim e^{-r/a}$ with
the radius $a=1.6$\AA was used to evaluate the form factors. Figure
\ref{fig:LDA-CD-woSO}(b)-(d) shows $D(\v k)$ overlaid with local OAM
when $\v k_F =2.27$\AA. The next set of figures in (e) and (f) is
obtained with $\v k_F = 0$\AA for $E_2$ and $E_3$ bands. Incident
photon direction is chosen $\hat{k}_\mathrm{ph} = (\cos 60^\circ, 0,
-\sin 60^\circ )$ in both sets. The photon energy dependence of the
NCD is obvious by comparing Fig. \ref{fig:LDA-CD-woSO}(c) with (e).
LDA calculation for light-element $d$-orbital bands are under way and
will be reported elsewhere, along with some experimental
results\cite{yonsei}. We comment that a potential modification of the
NCD formula may be necessary for heavy elements due to significant
spin-orbit interaction in the Hamiltonian and the consequent
modification of the current (momentum) operator coupling to $\v
A$\cite{gedik}. Our present results are, however, free from such
complications as we exclusively focus on spin-degenerate band
structures. Non-trivial OAM and NCD features are only due to the
degenerate multiple-orbital nature of the eigenstates.

\acknowledgments This work is supported by Mid-career Researcher
Program No. 2011-0015631 (JHH).

\end{document}